\documentclass[12pt]{article}

\usepackage[pdftex]{graphicx}

\usepackage{epsfig,cite}

\usepackage {amsmath}

\usepackage{color}
\usepackage{rotating}
\usepackage{amssymb}

\usepackage{slashed}

\include{epsf}

\newcount\timecount

\newcount\hours \newcount\minutes  \newcount\temp \newcount\pmhours

\hours = \time

\divide\hours by 60

\temp = \hours

\multiply\temp by 60

\minutes = \time

\advance\minutes by -\temp

\def\hour{\the\hours}

\def\minute{\ifnum\minutes<10 0\the\minutes

            \else\the\minutes\fi}

\def\clock{

\ifnum\hours=0 12:\minute\ AM

\else\ifnum\hours<12 \hour:\minute\ AM

      \else\ifnum\hours=12 12:\minute\ PM

            \else\ifnum\hours>12

                 \pmhours=\hours

                 \advance\pmhours by -12

                 \the\pmhours:\minute\ PM

                 \fi

            \fi

      \fi

\fi

}

\def\monthname{\relax\ifcase\month 0/\or January\or February\or

   March\or April\or May\or June\or July\or August\or September\or

   October\or November\or December\else\number\month/\fi}

\def\bold#1{\setbox0=\hbox{$#1$}%

     \kern-.025em\copy0\kern-\wd0

     \kern.05em\copy0\kern-\wd0

     \kern-.025em\raise.0433em\box0 }

\def\beq{\begin{equation}}

\def\eeq{\end{equation}}

\def\ga{\mathrel{\raise.3ex\hbox{$>$\kern-.75em\lower1ex\hbox{$\sim$}}}}

\def\la{\mathrel{\raise.3ex\hbox{$<$\kern-.75em\lower1ex\hbox{$\sim$}}}}

\def\gev{{\rm \, Ge\kern-0.125em V}}

\def\tev{{\rm \, Te\kern-0.125em V}}

\def\gyr{{\rm \, G\kern-0.125em yr}}

\def\gappeq{\mathrel{\rlap {\raise.5ex\hbox{$>$}}

{\lower.5ex\hbox{$\sim$}}}}

\def\lappeq{\mathrel{\rlap{\raise.5ex\hbox{$<$}}

{\lower.5ex\hbox{$\sim$}}}}

\def\Toprel#1\over#2{\mathrel{\mathop{#2}\limits^{#1}}}

\def\m12{m_{1\!/2}}

\def\bea{\begin{eqnarray}}

\def\eea{\end{eqnarray}}

\def\beqar{\begin{eqnarray}}

\def\eeqar{\end{eqnarray}}

\def\m{{\cal m}}

\begin{document}

\begin{titlepage}

\pagestyle{empty}

\baselineskip=21pt


\rightline{KCL-PH-TH/2012-28, LCTS/2012-14, CERN-PH-TH/2012-192}

\vskip 1in

\begin{center}

{\large {\bf Global Analysis of the Higgs Candidate with Mass $\sim 125$~GeV}}

\end{center}

\begin{center}

\vskip 0.2in

 {\bf John~Ellis}$^{1,2}$
and {\bf Tevong~You}$^{2,3}$

\vskip 0.1in

{\small {\it

$^1${Theoretical Particle Physics and Cosmology Group, Physics Department, \\
King's College London, London WC2R 2LS, UK}\\

$^2${TH Division, Physics Department, CERN, CH-1211 Geneva 23, Switzerland}\\

$^3${High Energy Physics Group, Blackett Laboratory, Imperial College, Prince Consort Road, London SW7 2AZ, UK}\\
}}

\vskip 0.2in

{\bf Abstract}

\end{center}

\baselineskip=18pt \noindent


{
We analyze the properties of the Higgs candidate
with mass $\sim 125$~GeV discovered by the CMS and ATLAS
Collaborations, constraining the possible deviations of its couplings from those
of a Standard Model Higgs boson. The CMS, ATLAS and Tevatron data are compatible with Standard
Model couplings to massive gauge bosons and fermions, and disfavour several
types of composite Higgs models unless their couplings resemble those in
the Standard Model. We show that the couplings of the Higgs candidate
are consistent with a linear dependence on particle masses, scaled by
the electroweak scale $v \sim 246$~GeV, the power law and the mass
scale both having uncertainties $\sim 20$\%.}


\vfill

\leftline{
July 2012}

\end{titlepage}

\baselineskip=18pt


\section{Introduction and Summary}

The hint of a possible new particle $h$ with mass $\sim 125$~GeV reported earlier by the
LHC experiments ATLAS and CMS~\cite{LHCcombined,ZZsearch, bbbarsearch, CMSdiphotonsearch, ATLASdiphotonsearch, ATLASWWsearch, CMSWWsearch, ATLAStautausearch, CMStautaumumu, CMSWHsearch,CMSprevdiphotonsearch},
has now become an indisputable discovery~\cite{ATLASICHEP2012,CMSICHEP2012, CMS8Combination, CMS8diphotonsearch, CMS8ZZsearch, CMS8WWsearch, CMS8tautausearch, CMS8bbsearch, ATLAS8ZZsearch, ATLAS8diphotonsearch,ATLAS8combination}, which has been supported by
new analyses from the Tevatron collider experiments CDF and D0 \cite{TevatronJulySearch}. There is a general
expectation that $h$ may be the long-sought Higgs boson~\cite{Higgsboson}, but many consistency checks must be made
before this identification can be confirmed. For example, it will be necessary to verify that the spin of the
$h$ particle is zero~\cite{JEDSHspin2} - the assignment assumed in searches in the $W W^*$ and $Z Z^*$ channels,
which is also consistent with the observation of $h$ decay into
$\gamma \gamma$ - and one would like to verify that the couplings of the $h$ to other
particles are proportional to their masses. Moreover, even if the $h$ particle passes these tests, other
measurements and consistency checks will be needed to see whether it stands alone or is the first representative
of a more complicated, possibly composite, electroweak symmetry-breaking sector.

Assuming that the $h$ particle does indeed have spin zero,
in this paper we explore the extent to which its couplings are constrained by the available data,
studying what limits can already be set on possible deviations from those of a Standard Model Higgs boson~\cite{mh125papers,postdiscovery}.
We treat as independent parameters the strengths of the $h$ couplings to massive vector bosons and to
different fermion species, including their indirect effects on loop-induced couplings to photon and gluon pairs
and assuming that the latter receive no significant contributions from particles beyond the Standard Model.

As reviewed below, one may parametrize the possible coupling deviations by coefficients $a_V$ and $c_f$
for vector bosons and fermions, respectively~\cite{ac}. One possibility is that these coefficients are universal, i.e.,
$a_W = a_Z \equiv a$ and $c_t = c_b = c_\tau = c_c = \dots \equiv c$, with the Standard Model corresponding
to $a = c = 1$. There has been some speculation that custodial symmetry might be broken with $a_W \ne a_Z$~\cite{dyszphilia},
and that couplings to some fermion species might be enhanced or suppressed. The present data are
insufficient to probe these possibilities very precisely, and the overall quality of our global fit, presented below,
indicates no need currently to adopt such hypotheses.

As already mentioned, a key prediction for the Standard Model Higgs boson is that its couplings to other
particles are proportional to their masses. We probe this issue here by considering purely phenomenological
parametrizations of the $h$ couplings of the forms $a_V = v (M_V^{2 \epsilon}/M^{(1 + 2 \epsilon)})$ and
$c_f = v ({m_f}^\epsilon/M^{1 + \epsilon})$, where for a Standard Model Higgs boson $\epsilon = 0$ and
 $M = v = 246$~GeV, the canonical Higgs vacuum expectation value (vev), corresponding to $a = c = 1$.

\begin{figure}
\vskip 0.5in
\begin{minipage}{8in}
\hspace*{-0.7in}
\centerline{\includegraphics[height=8cm]{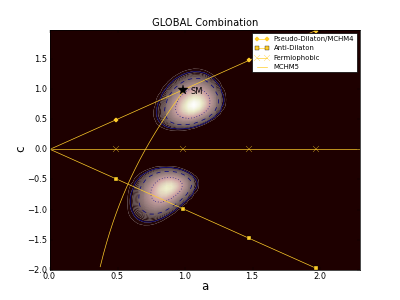}}
\hfill
\end{minipage}
\caption{
{\it
The constraints on the couplings $(a, c)$ of  the Higgs candidate $h$ with mass $\sim 125$~GeV
obtained from our global analysis of the available CMS, ATLAS, CDF and D0 data. The Standard Model
is represented by a black star, and the yellow lines represent various composite Higgs models
described in the text, which are disfavoured if they deviate strongly from the Standard Model.}} 
\label{fig:ac2}
\end{figure}

\begin{figure}
\vskip 0.5in
\vspace*{-0.75in}
\begin{minipage}{8in}
\hspace*{-0.7in}
\centerline{
{\includegraphics[height=6cm]{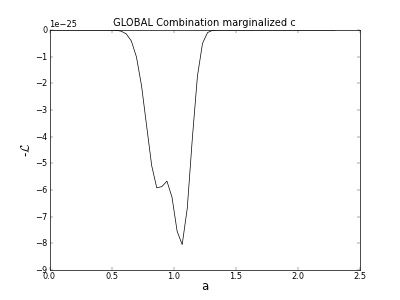}}
{\includegraphics[height=6cm]{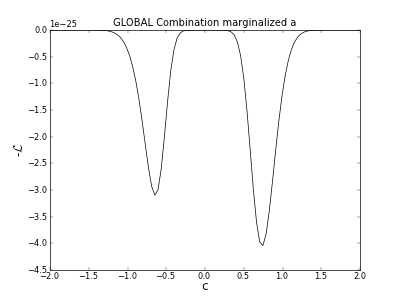}}
}
\hfill
\end{minipage}
\caption{
{\it
Marginalized one-dimensional projections on (left) the $a$ and (right) the $c$ axes of the likelihood function
for our global fit result shown in Fig~\protect\ref{fig:ac2}.  
}}  
\label{fig:ac1} 
\end{figure}

\begin{figure}
\vskip 0.5in
\vspace*{-0.75in}
\begin{minipage}{8in}
\hspace*{-0.7in}
\centerline{\includegraphics[height=8cm]{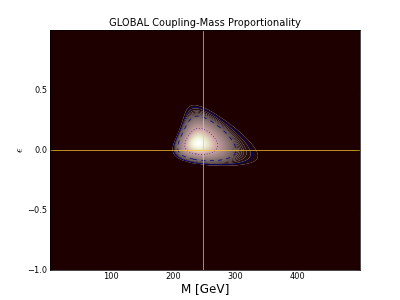}}
\hfill
\end{minipage}
\caption{
{\it
The constraints on the scaling parameters $(\epsilon, M)$ of  the Higgs candidate $h$ with mass $\sim 125$~GeV
obtained from our global analysis of the available CMS, ATLAS, CDF and D0 data. The Standard Model
corresponds to the intersection of the yellow cross-hairs. The data are close to the `bull's eye'.
}} 
\label{fig:epsilonM2} 
\end{figure}

\begin{figure}
\vskip 0.5in
\vspace*{-0.75in}
\begin{minipage}{8in}
\hspace*{-0.7in}
\centerline{{\includegraphics[height=6cm]{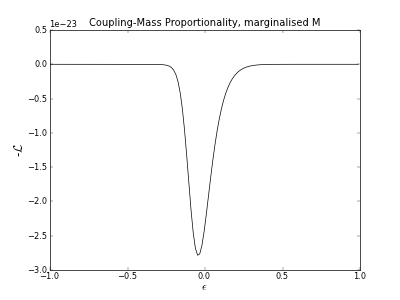}}
{\includegraphics[height=6cm]{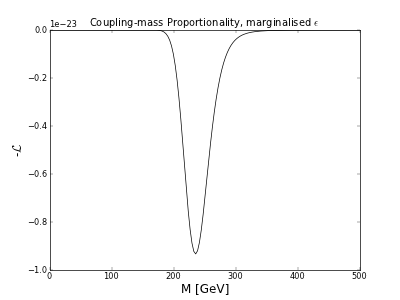}}}
\hfill
\end{minipage}
\caption{
{\it
Marginalized one-dimensional projections on (left) the $\epsilon$ and (right) $M$ axes of the likelihood function
for our global fit result shown in Fig~\protect\ref{fig:epsilonM2}.  
}}  
\label{fig:epsilonM1} 
\end{figure}

Figures~\ref{fig:ac2}, \ref{fig:ac1}, \ref{fig:epsilonM2} and \ref{fig:epsilonM1} display our main results. They
quantify the extent to which the $h$ particle walks like a Higgs and quacks like a Higgs.

Fig.~\ref{fig:ac2} shows the result in the $(a, c)$
plane of our global fit to data on the $h$ couplings from the Tevatron experiments and from the combined 7 and
8-TeV event samples of ATLAS and CMS. We see reasonable consistency with the Standard Model prediction:
the overall best-fit region has $c > 0$ and, whilst the best fit has $a > 1$ and $c < 1$ (see also the marginalized
one-dimensional likelihoods of our fit result projected on the
$a$ and $c$ axes shown in Fig~\ref{fig:ac1}), the Standard Model prediction lies within the 68\% CL region. As we discuss in more detail below,
Figs.~\ref{fig:ac2} and \ref{fig:ac1} impose important constraints on composite Higgs models, disfavouring several
such models unless their predictions resemble those of the Standard Model.

\begin{figure}
\vskip 0.5in
\vspace*{-0.75in}
\begin{minipage}{8in}
\hspace*{-0.7in}
\centerline{\includegraphics[height=8cm]{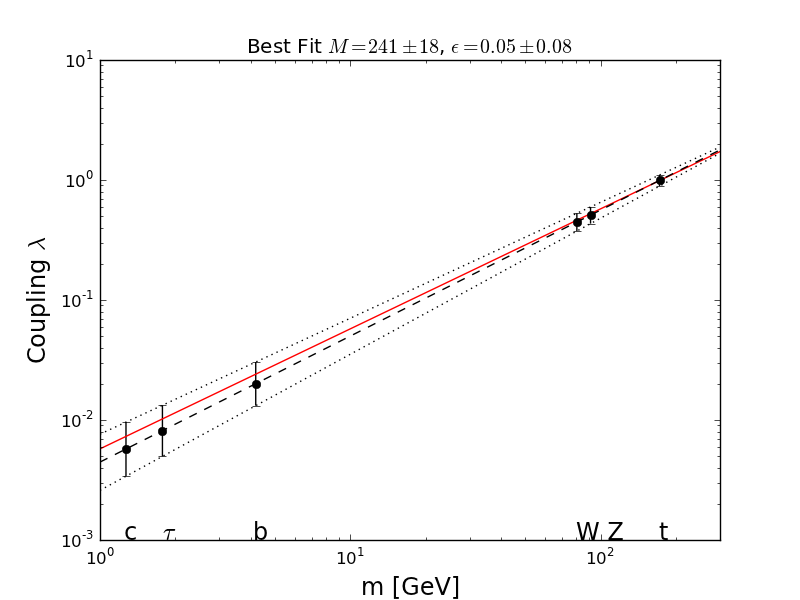}}
\hfill
\end{minipage}
\caption{
{\it
The mass dependence of the $h$ couplings found in our $(\epsilon, M)$ fit. The vertical error bars correspond to the
uncertainties shown in Fig.~\protect\ref{fig:epsilonM1}. The dashed line is our best fit, and the dotted lines correspond
to $\pm 1 \sigma$ variations in $(\epsilon, M)$. The Standard Model prediction
that Higgs couplings should be proportional to the masses of other particles with $M = v$, shown by the diagonal solid red line, is
completely consistent with the data.
}} 
\label{fig:Mdep} 
\end{figure}

Fig.~\ref{fig:epsilonM2} displays the result of our global fit in
the $(\epsilon, M)$ plane, where we see excellent consistency with the Higgs hypothesis: $M = v, \epsilon = 0$.
This is also seen in Fig.~\ref{fig:epsilonM1}, which displays the marginalized one-dimensional likelihood projections of our fit result on the
$M$ and $\epsilon$ axes. The couplings of the $h$ particle are clearly inconsistent with any mass-independent
scenario, which would require $\epsilon = -1$. Fig.~\ref{fig:Mdep} provides another way of understanding this observation. The points with vertical error bars
represent the values of the $h$ couplings to different particles found in our global fit to the parameters $(\epsilon, M)$.
The diagonal dashed line is our best fit to $(\epsilon, M)$ and the dotted lines are given by the $\pm 1 \sigma$
ranges in these parameters, as given in the upper legend of the plot. The solid red line in Fig.~\ref{fig:Mdep} represents the Standard Model
prediction (\ref{SMcouplings}), which is compatible within errors with the measurements, as already discussed.

In subsequent Sections we describe how these results were obtained, and
present more details of our analysis. In Section~2 we review the phenomenological
frameworks we employ, and in Section~3 we describe our calculational procedure, which follows
closely that in~\cite{EY}. In Section~4 we describe the data set we use, focusing in particular on the
recent update from the Tevatron experiments as well as the
recent preliminary results from $\sim 5$/fb of 8-TeV LHC data in each of ATLAS and CMS. In Section~5 we present in more detail our results in
the $(a, c)$ plane, discussing their implications for pseudo-dilatons~\cite{pseudoDG,pseudoDG2} and other 
composite Higgs scenarios~\cite{compositeHiggs}~\footnote{Radion models~\cite{radion} are closely related.},
as well as fermiophobic~\cite{fermiophobic} and gaugephobic~\cite{gaugephobic} models. In Section~6 we discuss in more detail our results in
the $(\epsilon, M)$ plane,  and in Section~7 we present our conclusions and discuss
the prospects that future data may soon clarify further the nature of the $\sim 125$~GeV Higgs candidate $h$.

\section{Phenomenological Framework}

We use the following nonlinear low-energy effective Lagrangian for the
electroweak symmetry-breaking sector~\cite{ac,pseudoDG}, see also~\cite{pionL,dilatonL}:
\bea
{\cal L}_{eff} \; & = & \; \frac{v^2}{4} {\rm Tr} \left(D_\mu U D^\mu U^\dagger \right) \times \left[ 1 + 2 a \frac{h}{v} +  \dots \right] \nonumber \\
& - & \frac{v}{\sqrt{2}} \Sigma_f {\bar f}_L\lambda_f f_R \left[ 1 + c_f \frac{h}{v} + \dots \right] + h.c.
\label{effL}
\eea
where $U$ is a unitary $2 \times 2$ matrix parametrizing the three Nambu-Goldstone fields
that are `eaten' by the $W^\pm$ and $Z^0$, giving them masses, $v \sim 246$~GeV is the
conventional electroweak symmetry-breaking scale, $h$ is a field describing the Higgs candidate with mass
$\sim 125$~GeV discovered by ATLAS and CMS, and $\lambda_f$ is the Yukawa coupling
of the fermion flavour $f$ in the Standard Model. The coefficients $a$ and $c_f$ parametrize the 
deviations of the $h$ couplings to massive
vector bosons and to fermions, respectively, from those of the Higgs boson in the Standard Model.
In writing (\ref{effL}), we have implicitly assumed a custodial symmetry: $a_W = a_Z = a$,
an assumption whose plausibility can be judged from the overall quality of our fit.

Also relevant for the phenomenology of the Higgs candidate $h$ are
its dimension-5 loop-induced couplings to $gg$ and $\gamma \gamma$~\cite{EGN,traceanomaly}:
\begin{equation}
{\cal L}_{\Delta} \; = \; - \left[ \frac{\alpha_s}{8 \pi} b_s 
G_{a \mu \nu} G_a^{\mu \nu} + \frac{\alpha_{em}}{8 \pi} b_{em} F_{\mu \nu} F^{\mu \nu} \right] \left(\frac{h}{V}\right) .
\label{triangles}
\end{equation}
We assume here that, as in the Standard Model, only the top quark makes a significant contribution to the coefficient $b_s$, 
and only the top quark and the $W^\pm$ contribute significantly to $b_{em}$ (with opposite signs in the Standard Model~\cite{EGN}).

We recall that, in a scenario in which $h$ is associated with a pseudo-dilaton field $\chi$ with vev $V$, one has
\beq
a \; = \; c \; = \frac{v}{V} .
\label{dilatonratio}
\eeq
One may also consider
scenarios in which $h$ is a pseudo-Goldstone boson of some
higher-order chiral symmetry that is broken down to the SU(2) $\times$ SU(2) of the Standard Model
Higgs sector. Among such composite models with an SO(5)/SO(4) structure \cite{minimalCompositeHiggs},
one may consider the MCHM4 option - in which the Standard Model fermions 
are embedded in spinorial representations of SO(5) and
\beq
a \; = \; c \; = \; \sqrt{1 - \xi} ,
\label{MCHM4}
\eeq
where $\xi \equiv (v/f)^2$ with $f$ a compositeness scale (which is equivalent to the pseudo-dilaton
model with $v/V \to \sqrt{1 - \xi}$), or the MCHM5 option - in which the Standard Model fermions 
are embedded in fundamental representations of SO(5) and
\beq
a \; = \; \sqrt{1 - \xi} \; , \; c \; = \; \frac{1 - 2 \xi}{\sqrt{1 - \xi}} .
\label{MCHM5}
\eeq
As discussed in~\cite{EY},
this interpolates between the Standard Model (obtained in the limit $\xi \to 0$), a specific fermiophobic scenario with $a = \sqrt{3}/2$
(obtained in the limit $\xi \to 1/2$), an `anti-dilaton' model with $a = - c = 1/\sqrt{3}$ (obtained when $\xi = 2/3$), and a
gaugephobic model (obtained when $\xi \to 1$).

In addition to these theoretically-motivated models, we also consider the completely phenomenological
possibility that $h$ couples to other particles proportionally to some power of their masses. Thus, we
generalize the Standard Model couplings
\beq
\lambda_f \; = \; \sqrt{2} \frac{m_f}{v}, \; g_V \; = \; 2 \frac{m_V^2}{v}
\label{SMcouplings}
\eeq
to the following forms of couplings with anomalous scaling laws:
\beq
\lambda_f \; = \; \sqrt{2} \left(\frac{m_f}{M}\right)^{1 + \epsilon}, \; g_V \; = \; 2 \left(\frac{m_V^{2(1 + \epsilon)}}{M^{1 + 2\epsilon}}\right) .
\label{scalingcouplings}
\eeq
The Standard Model is recovered in the double limit $\epsilon \to 0, M \to v$, whereas the
pseudo-dilaton/MCHM4 scenario would correspond to $\epsilon = 0$ and $M = V \ne v$, in general.
In terms of the parameterization (\ref{effL}), the parametrizations (\ref{scalingcouplings})
correspond to
\beq
c_f \; = \; v \left(\frac{{m_f}^\epsilon}{M^{1 + \epsilon}}\right), \; a_V \; = \; v \left(\frac{M_V^{2 \epsilon}}{M^{(1 + 2 \epsilon)}}\right) .
\label{scalingca}
\eeq
After presenting our global fits to the parameters $a, c$, we shall explore the extent to which the
data already indicate that $h$ couples to other particles proportionally to masses, i.e., with $\epsilon =0$, and a normalization $M$
similar to $v =246$~GeV.

\section{Calculational Procedure}

Assuming that the Higgs candidate $h$ has no non-standard production or decay modes,
its production cross-sections and decay widths are related to those of the Standard Model Higgs boson
by simple factors of $a$ and $c$.
Assuming that gluon-gluon fusion (gg) and vector-boson fusion (VBF) dominate over the other processes, as at the LHC,
one may combine their respective production rescaling factors $R\equiv\sigma/\sigma_{\text{SM}}$
and cut efficiencies $\xi_{\text{gg,VBF}}$ to obtain a total production rescaling factor 
\begin{equation}
	R_{\text{prod}} = \frac{\xi_{gg}F_{gg}R_{gg} + \xi_{\text{VBF}}(1-F_{gg})R_{\text{VBF}}}{\xi_{gg}F_{gg} + \xi_{\text{VBF}}(1-F_{gg})}		\quad ,
\label{production}
\end{equation}
where $F_{gg} \equiv \sigma^{\text{SM}}_{gg}/\sigma^\text{SM}_\text{tot}$. In the case of the Tevatron, where associated production (AP)
is more important that VBF, one may use (\ref{production}) with the replacement VBF $\to$ AP throughout. For the CMS diphoton subchannels, the collaboration provides a full breakdown of the percentage contribution from all production mechanisms which can be used directly instead of the $\xi_i F_i$ factors above.

Similarly, relative to the Standard Model predictions,
the decay widths $R\equiv\Gamma/\Gamma_{\text{SM}}$ to massive vector bosons, fermions and photons are given, respectively, by
\begin{equation}
	R_{VV} = a^2	\quad , \quad	R_{\bar{f}f} = c^2	\quad , \quad	R_{\gamma\gamma} = \frac{(-\frac{8}{3}cF_t + aF_w)^2}{(-\frac{8}{3}F_t+F_w)^2}	\quad , 
\label{loopfactors}
\end{equation}
where the loop factors $F_{t,w}$ were given, e.g., in~\cite{EY}.
The principal dependences of the
different Higgs-like signals on the rescaling factors $(a, c)$ are summarized in Table~\ref{table:channeleffparamssensitivity},
which is adapted from~\cite{EY}. It is important to emphasize that, since production mechanisms are in general also sensitive to
both $a$ and $c$, as well as decay branching ratios, their dependences also provide important constraints on model
parameters.

\begin{table}[h!]
	\center
	\begin{tabular}{ | c | c | c | c | c |}
		\hline
		 & \multicolumn{2}{|c|}{Production sensitive to} & \multicolumn{2}{|c|}{Decay sensitive to} \\ 
		channel & \quad $a$ \quad & \quad $c$ \quad & \quad $a$ \quad & \quad $c$ \quad  \\ \hline
		$\gamma\gamma$ & $\checkmark$ & $\checkmark$ & $\checkmark$ & $\checkmark$ \\ \hline
		$\gamma\gamma$ VBF & $\checkmark$ & $\times$ & $\checkmark$ & $\checkmark$ \\ \hline
		WW & $\checkmark$ & $\checkmark$ & $\checkmark$ & $\times$ \\ \hline 
		WW 2-jet & $\checkmark$ & $\times$ & $\checkmark$ & $\times$ \\ \hline
		WW 0,1-jet & $\times$ & $\checkmark$ & $\checkmark$ & $\times$ \\ \hline
		$b\bar{b}$ (VH) & $\checkmark$ & $\times$ & $\times$ & $\checkmark$  \\ \hline
		$b\bar{b}$ ($\bar{t}tH$) & $\times$ & $\checkmark$ & $\times$ & $\checkmark$  \\ \hline
		ZZ & $\checkmark$ & $\checkmark$ & $\checkmark$ & $\times$ \\ \hline
		$\tau\tau$ & $\checkmark$ & $\checkmark$ & $\times$ & $\checkmark$ \\ \hline 
		$\tau\tau$ (VBF, VH) & $\checkmark$ & $\times$ & $\times$ & $\checkmark$ \\ \hline 
	\end{tabular}
	\caption{\it The dominant dependences on the model parameters $(a, c)$ (\protect\ref{effL}) of the $h$ detection and search
	channels discussed	in this paper, adapted from~\protect\cite{EY}.}
	\label{table:channeleffparamssensitivity}
\end{table}

The signal strength modification factor $\mu^i \equiv n^i_s / (n^i_s)^\text{SM}$ in any given channel $i$ is the product
of the production and decay rescalings: $R \equiv R^i_{\text{prod}}\cdot (R^i_\text{decay}/R_\text{tot.})$. 
In the absence of more detailed experimental information, we follow~\cite{Contino} as in~\cite{EY}, assuming
that in each channel the underlying likelihood $p(n_\text{obs} | \mu n_s^\text{SM} + n_b)$ obeys a Poisson distribution,
and use the approximation $\sigma_{\text{obs}} \simeq \sigma_{\text{exp}} = \mu^{95\%}_\text{exp}/1.96$  for the standard deviation
to solve for the central value $\bar{\mu}$ in the equation:
\begin{equation}
	\frac{\int^{\mu^{95\%_\text{obs}}}_0 e^{-\frac{(\mu - \bar{\mu})^2}{2\sigma^2_\text{obs}}} d\mu}{\int^\infty_0 e^{-\frac{(\mu - \bar{\mu})^2}{2\sigma^2_\text{obs}}} d\mu} = 0.95	\quad .
\end{equation} 
%
The posterior probability density function is then given by 
\begin{equation}
	p(\mu | n_\text{obs}) = p(n_\text{obs} | \mu n_s^\text{SM} + n_b)\cdot\pi(\mu) \approx \frac{1}{\sqrt{2\pi\sigma_\text{obs}^2}}e^{-\frac{(\mu-\bar{\mu})^2}{2\sigma_\text{obs}^2}}	\quad ,
\end{equation}
with $\pi(\mu)$ generally assumed {\it a priori} to be flat within the range of interest and zero outside.

\section{Experimental Data Set} 

We use the latest available information from $\sim 5$/fb of LHC data obtained at each of 7 and 8 TeV in the centre of mass presented at ICHEP 2012~\cite{CMSICHEP2012,ATLASICHEP2012}, and $\sim 10$/fb of Tevatron data analyzed in \cite{TevatronJulySearch}. In addition to~\cite{CMSICHEP2012}, 
the CMS Collaboration provides additional information on its 7 and 8 TeV fits separately in public analysis notes, see below.

\begin{enumerate}
	\item The CMS and ATLAS searches in the channel $h \to ZZ \to 4 \ell^\pm$ are treated as inclusive for both 7 and 8 
	TeV~\cite{ZZsearch, ATLAS8ZZsearch, CMS8Combination}.
	\item The searches in the $h \to \bar{b}b$ VH channel are assumed to be dominated by associated production, with the Tevatron data updated from \cite{EY} to the latest results in \cite{TevatronJulySearch}, the 7 TeV Moriond results are used for ATLAS~\cite{bbbarsearch}, and the CMS 7 and 8 TeV fits are obtained from \cite{CMS8Combination}. In addition the CMS 7 TeV ${\bar t}tH$ channel is included from \cite{CMS8bbsearch}.
	\item The diphoton likelihoods in the ATLAS searches at 7 TeV were obtained from \cite{ATLASdiphotonsearch} as explained in \cite{EY}. In \cite{CMS8diphotonsearch} CMS provides central values and one sigma error bars for both 7 and 8 TeV searches in four inclusive sub-channels dominated by gluon fusion and one or two di-jet categories. We treat the 8 TeV ATLAS results inclusively since the sub-channel best fit values are only provided for $m_h = 126.5$~\cite{ATLAS8diphotonsearch}~\footnote{As discussed below, our results are quite insensitive to the assumed for $h$ in the range $[124, 127]$~GeV.}. The Tevatron search from \cite{TevatronJulySearch}  are also included.  
	\item The Tevatron results for $h \to W^+W^-$ are updated from \cite{TevatronJulySearch}. The corresponding ATLAS results for 7 TeV \cite{ATLASWWsearch} are supplemented by the 8 TeV data made public recently in \cite{ATLAS8WWsearch}. CMS provide fits in the 0,1 and 2-jet categories for both 7 and 8 TeV~\cite{CMS8WWsearch, CMS8Combination}.   
	\item The ATLAS $\tau\tau$ searches at 7 TeV are treated as inclusive~\cite{ATLAStautausearch}. For CMS we use the best fits provided for the 0,1-jet and VBF channels at 7 and 8 TeV in \cite{CMS8tautausearch, CMS8Combination}. At 7 TeV there is also an additional CMS search in the VH channel.
\end{enumerate}

As mentioned in the previous section, we use in our fit the CMS information on the percentage contribution from each production mechanism for all the diphoton sub-categories at 7 and 8 TeV separately. We treated the $\tau\tau$ VBF categories assuming $\sim$30\% contamination from gluon fusion in the production mechanism. As mentioned in \cite{EY}, we expect that
our global analysis is only accurate to $\sim$20\% due to the limited experimental information
available so far~\cite{HouchesRecommendations}.

\section{Results}

\subsection{Tevatron data}

\begin{figure}
\vskip 0.5in
\vspace*{-0.75in}
\begin{minipage}{8in}
\includegraphics[height=6cm]{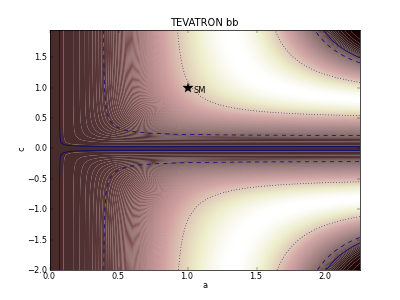}
\includegraphics[height=6cm]{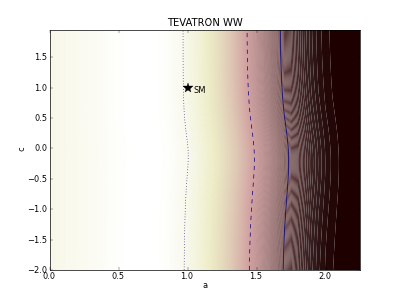}
\end{minipage}
\begin{minipage}{8in}
\includegraphics[height=6cm]{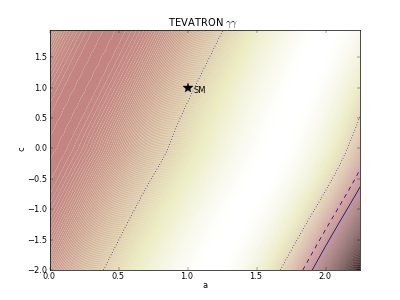}
\includegraphics[height=6cm]{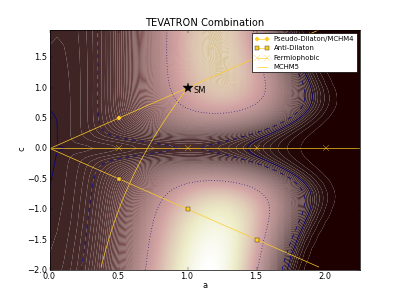}
\hfill
\end{minipage}
\caption{
{\it
Constraints on the couplings $(a, c)$ of the Higgs candidate $h$ with mass $\sim 125$~GeV
arising from the CDF and D0 data on (upper left) ${\bar b} b$, (upper right) $\tau^+ \tau^-$,
and (lower left) $\gamma \gamma$ final states. 
The lower right panel displays the combination of these ICHEP 2012 CMS constraints, together with lines
representing the pseudo-dilaton/MCHM4, anti-dilaton, fermiophobic and MCHM5 scenarios.
In these and subsequent analogous plots,
the most likely regions have the lightest shading,
the dotted lines are 68\% CL contours, the dashed lines are 95\% CL contours, and the solid lines
are 99\% CL contours.}}
\label{fig:Tevatron} 
\end{figure}

We consider first the fit to the recent Tevatron data in terms of $(a, c)$ that is
shown in Fig.~\ref{fig:Tevatron}. We recall that the Tevatron experiments CDF and D0
provide information on the associated production (AP) of $h$ followed by its decay
into ${\bar b} b$ (upper left panel of Fig.~\ref{fig:Tevatron}), as well as inclusive measurements
of $h \to W W^*$ decay (upper right panel) and now also $h \to \gamma \gamma$ decay
(lower left panel). The central value of the $h \to {\bar b} b$ signal is somewhat stronger
than expected in the Standard Model, disfavouring fermiophobic models and
corresponding to the possibility that either $a$ and/or $c > 1$.
However, the Tevatron $h \to WW^*$ signal is relatively weak, disfavouring large $a$. The
$h \to \gamma \gamma$ signal is relatively strong, but very uncertain. In combination
(lower right panel of Fig.~\ref{fig:Tevatron}), the Tevatron data are compatible with the
Standard Model, while favouring slightly $a, c > 1$.

\subsection{CMS data}

\begin{figure}
\vskip 0.5in
\vspace*{-1.1in}
\begin{minipage}{8in}
\includegraphics[height=6cm]{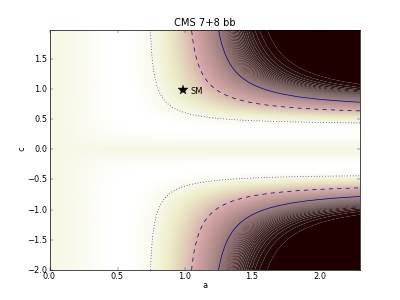}
\hspace*{0.2in}
\includegraphics[height=6cm]{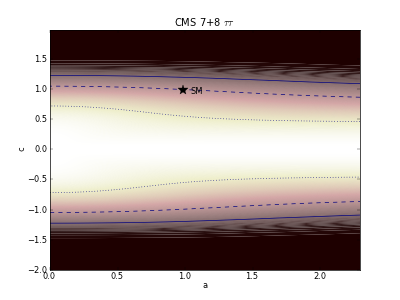}
\hfill
\end{minipage}
\begin{minipage}{8in}
\includegraphics[height=6cm]{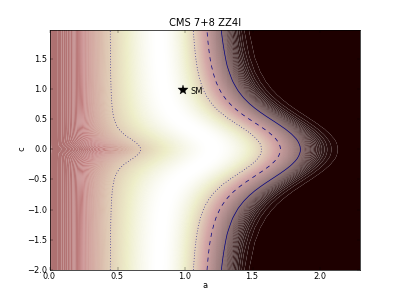}
\hspace*{0.2in}
\includegraphics[height=6cm]{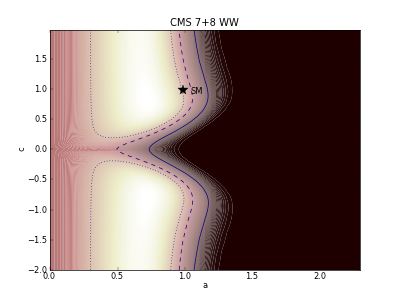}
\hfill
\end{minipage}
\begin{minipage}{8in}
\includegraphics[height=6cm]{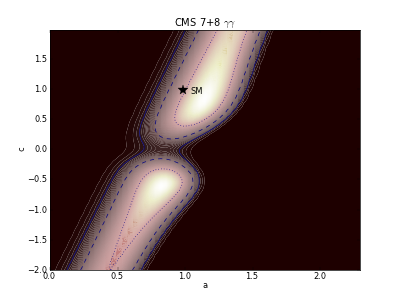}
\hspace*{0.2in}
\includegraphics[height=6cm]{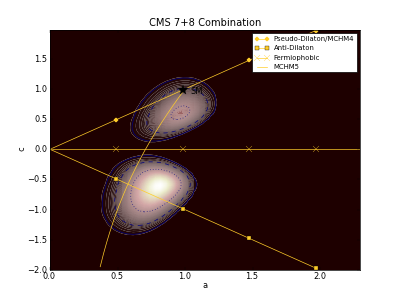}
\hfill
\end{minipage}
\caption{
{\it
Constraints on the couplings $(a, c)$ of the Higgs candidate $h$ with mass $\sim 125$~GeV
arising from the ICHEP 2012 CMS data on (top left) ${\bar b} b$, (top right) $\tau^+ \tau^-$,
(centre left)  $Z Z^*$, (centre right) $W W^*$ and (bottom left) $\gamma \gamma$ final states. 
The bottom right panel displays the combination of these ICHEP 2012 CMS constraints, together with lines
representing the pseudo-dilaton/MCHM4, anti-dilaton, fermiophobic and MCHM5 scenarios.
As in other analogous plots,
the most likely regions have the lightest shading,
the dotted lines are 68\% CL contours, the dashed lines are 95\% CL contours, and the solid lines
are 99\% CL contours.}} 
\label{fig:CMS} 
\end{figure}

We now turn to the analysis of the ICHEP 2012 CMS data shown in Fig.~\ref{fig:CMS}.
The $h \to {\bar b} b$ search (top left panel) was based on AP and ${\bar t} t$ Higgsstrahlung (HS)
event selections, the former being more sensitive. The overall signal strength
is somewhat below that expected in the Standard Model, slightly favouring $a, c < 1$,
but very compatible with the Standard Model. The $h \to \tau^+ \tau^-$ search (top right panel)
was based on a combination of event selections favouring gluon-gluon fusion, VBF and AP
production mechanisms. Once again the overall signal is weaker than expected in the
Standard Model, but not very significantly. The $h \to Z Z^*$ signal (middle left panel) has the strength
expected in the Standard Model, disfavouring $a \ll 1$. The $h \to W W^*$ search shown in the
middle right panel of Fig.~\ref{fig:CMS} was based on
a combination of event selections favouring gluon-gluon fusion and VBF production
mechanisms, and the deficit compared to the Standard Model is not very significant. Finally,
the $h \to \gamma \gamma$ event selection includes samples with and without enhanced
VBF contributions. As discussed in~\cite{EY}, since the $h \to \gamma \gamma$ decay
amplitude contains both $t$ and $W$ loops, which interfere, it provides unique discrimination
between the cases $a >$ and $< 0$, as seen in the bottom left panel of Fig.~\ref{fig:CMS}.
Although the $\gamma \gamma$ signal strength is somewhat stronger than in the Standard Model,
particularly in the VBF-enhanced sample, the discrepancy is not highly significant. Turning now to the
overall combination of CMS data shown in the bottom right panel of Fig.~\ref{fig:CMS}, we see that
the overall best fit is in a region with $c < 1$, driven by the $\gamma \gamma$ channel. 
The favoured region with $c > 0$ is compatible with the Standard Model, 
with $a \sim 1$ though $c < 1$ is somewhat favoured~\footnote{When we restrict our fit to $c > 0$,
we obtain a result very similar result to that reported by the CMS Collaboration~\cite{CMSICHEP2012}.}.
Comparing with the analogous panel in Fig.~1 of~\cite{EY}, we see that the accuracy in the
determination of the $h$ couplings has improved significantly.

\subsection{ATLAS data}

\begin{figure}
\vskip 0.5in
\vspace*{-1.1in}
\begin{minipage}{8in}
\includegraphics[height=6cm]{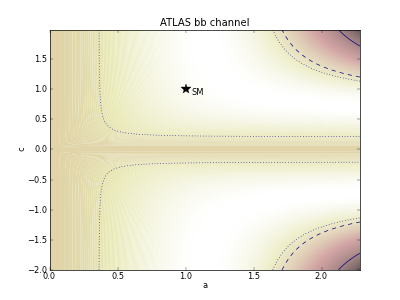}
\hspace*{0.2in}
\includegraphics[height=6cm]{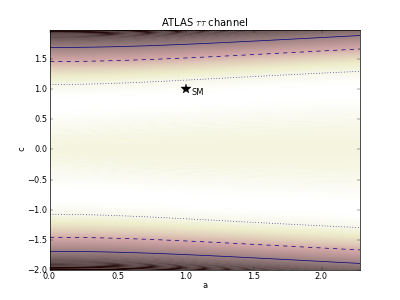}
\hfill
\end{minipage}
\begin{minipage}{8in}
\includegraphics[height=6cm]{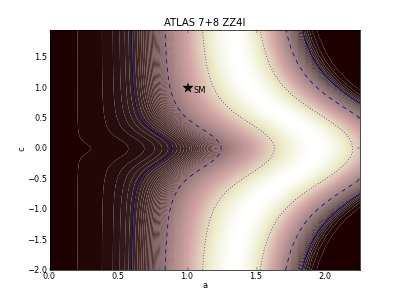}
\hspace*{0.2in}
\includegraphics[height=6cm]{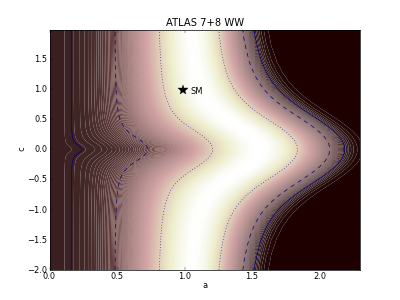}
\hfill
\end{minipage}
\begin{minipage}{8in}
\includegraphics[height=6cm]{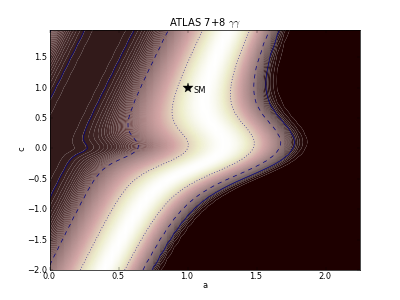}
\hspace*{0.2in}
\includegraphics[height=6cm]{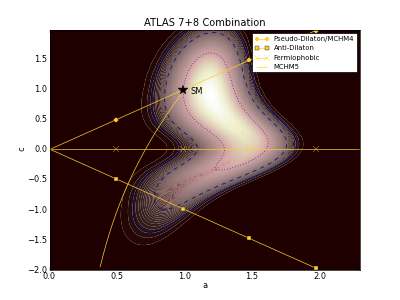}
\hspace*{0.2in}
\hfill
\end{minipage}
\caption{
{\it
Constraints on the couplings $(a, c)$ of the Higgs candidate $h$ with mass $\sim 125$~GeV
arising from the ATLAS data on (top left) ${\bar b} b$, (top right) $\tau^+ \tau^-$,
(centre left)  $Z Z^*$, (centre right) $W W^*$ and (bottom left) $\gamma \gamma$ final states. 
The bottom right panel displays the combination of these ICHEP 2012 ATLAS constraints, together with lines
representing the pseudo-dilaton/MCHM4, anti-dilaton, fermiophobic and MCHM5 scenarios.
}} 
\label{fig:ATLAS} 
\end{figure}

We now present a similar analysis of the available ATLAS data, which yields
the results shown in Fig.~\ref{fig:ATLAS}. As in the previous figure, the top left panel
displays the constraint in the $(a, c)$ plane provided by the $h \to {\bar b} b$
search, which in the ATLAS case is based on $\sim 5$/fb of data at 7 TeV, as is the $h \to \tau^+ \tau^-$
constraint shown in the top right panel of Fig.~\ref{fig:ATLAS}. These panels are the same as the
corresponding panels in Fig.~5 of ~\cite{EY}. The middle left panel of
Fig.~\ref{fig:ATLAS} displays the $h \to Z Z^*$ constraint including also $\sim 5$/fb of data at 8 TeV:
we see that the central value of the signal strength lies somewhat above the value expected in
the Standard Model, corresponding to $a > 1$. The central value of the $h \to WW^*$
signal shown in the middle right panel of  Fig.~\ref{fig:ATLAS} (which is based on $\sim 5$/fb of data each at 7 TeV and 8 TeV)
has $a > 1$, but is consistent with $a = 1$ at the 68\% CL. Finally, the combined ATLAS 7- and 8-TeV search for $h \to \gamma \gamma$
shown in the bottom right panel of Fig.~\ref{fig:ATLAS} yields a central value of the strength lying somewhat
above the Standard Model value, which is reflected in the preferred region of the $(a, c)$ plane.
The overall combination of the available ATLAS constraints, shown in the bottom right panel,
indicates a general preference for $a > 0$, with values of $a > 1, c < 1$ being favoured.

\section{Combined Results in the $(a,c)$ Plane and Implications for Models}

Looking at the bottom right panels of Figs.~\ref{fig:Tevatron}, \ref{fig:CMS} and \ref{fig:ATLAS},
the general features of Fig.~\ref{fig:ac2} can now be understood. Since they do not have
high sensitivity to $h \to \gamma \gamma$, the Tevatron data are unable to discriminate
between the upper and lower halves of the $(a, c)$ plane. The CMS data do have high
sensitivity to $h \to \gamma \gamma$, leading to some asymmetry between the upper and lower
halves of the $(a, c)$ plane, with a preference for $c < 0$. On the other
hand, the ATLAS $h \to \gamma \gamma$ and $W W^*$ data provide some preference for $c > 0$.
Generally speaking, the Tevatron data prefer $a, c >1$, whereas the CMS and ATLAS data prefer $a > 1$ and $c < 1$. 
The overall result, shown in Fig.~\ref{fig:ac1}, is that the
global combination prefers $a > 1$ (left panel) and $c > 0$ (right panel), though not very significantly, and
the favoured region with $c > 0$ has $a$ slightly $> 1$ (left panel) and $c$ slightly $ < 1$ (right panel).

The absence of strong ${\bar b}b$ and $\tau^+ \tau^-$ signals at the LHC favours
speculation that $c  \ll 1$, but in models with a universal coefficient $c$ for all fermions, this
is not the whole story. The fact the total $h$ cross section is compatible with the Standard Model
indicates that the $h {\bar t} t$ couplings should be close to the Standard Model value, corresponding
to $c \sim 1$. Moreover, the Tevatron experiments report evidence for a strong ${\bar b}b$ decay signal, and
as this is the dominant decay mode in the Standard Model the
whole pattern of $h$ decays would be very different if $c \ll 1$.
The right panel of Fig.~\ref{fig:ac1} is the net result: no significant discrepancy with
the Standard Model if $c$ is universal. 

Likewise, the absence of a strong $h \to W W^*$ signal in the Tevatron, CMS and the ATLAS 7-TeV data
might have led one to speculate that $a < 1$, or
even that $a_W \ne a_Z$ with custodial symmetry broken~\cite{dyszphilia}. However, the fact that the $h \to \gamma \gamma$ signals reported by both
ATLAS and CMS are on the high side suggests that the $h \gamma \gamma$ loop amplitude in (\ref{loopfactors}) must
receive an important contribution from the $W^\pm$ loops, which should be dominant. This and the ATLAS 8-TeV data suggest that
$a_W$ cannot be very small, and favours $a \equiv a_W = a_Z$ not $\sim 1$ in our fit.

We display in Fig.~\ref{fig:ac2} yellow lines corresponding to the predictions of the pseudo-dilaton and MCHM4 models
(\ref{dilatonratio}, \ref{MCHM4})
($a = c = v/V, \sqrt{1 - \xi}$), `anti-dilaton' models ($a = -c$), the MCHM5 model (\ref{MCHM5}) ($a = \sqrt{1 - \xi}, c = (1 - 2 \xi/\sqrt{1 - \xi})$),
fermiophobic models ($c = 0$) and gaugephobic models ($a = 0$)~\footnote{See also the combination panels in
Figs.~\ref{fig:Tevatron}, \ref{fig:CMS} and \ref{fig:ATLAS} for the corresponding individual comparisons with
Tevatron, CMS and ATLAS data, respectively.}. We see in the right panel of Fig.~\ref{fig:epsilonM1} that models
with an overall scale $M$ similar to the value $v = 246$~GeV in the Standard Model are strongly favoured. These
correspond to the cases $V \sim v$ in pseudo-dilaton models and $\xi \sim 0$ in the MCHM4 model. However, in
pseudo-dilaton models, in particular, there may be additional heavy particles contributing to the loop coefficients
$b_s, b_{em}$ in (\ref{triangles})~\cite{pseudoDG2}, so this observation is model-dependent and further analysis is needed~\cite{newMY}. We also
see in Fig.~\ref{fig:ac2} that the $\xi \to 0$ limit of the MCHM5 model is preferred, while `anti-dilaton' models are
slightly disfavoured compared to pseudo-dilaton models, and would prefer $a = -c < 1$: see also the right panel
of Fig.~\ref{fig:ac1}. Finally, we observe that
the fermiophobic and gaugephobic models are strongly disfavoured.

\section{Combined Results in the $(\epsilon,M)$ Plane}

The $h$ particle is clearly very different from any other known `fundamental' particle. The fact that it decays into
$\gamma \gamma$ implies that it cannot have spin one, and hence cannot be a gauge boson. It may well have
spin zero: this remains to be demonstrated, though this hypothesis has been used in the $WW^*$ and $ZZ^*$
event selections. If it has spin 2, that would make it an even more remarkable discovery. If it does have spin zero,
there is no reason why its couplings to different fermion generations (for example) should be universal, and the
Higgs hypothesis suggests that its couplings to other particles should be proportional to their masses.

We now discuss the light on the nature of the $h$ particle that is cast by Figs.~\ref{fig:epsilonM2}, \ref{fig:epsilonM1} and \ref{fig:Mdep}.
In particular, Fig.~\ref{fig:epsilonM2} suggests that the data are heading straight towards the Higgs `bull's eye'
at the cross-hairs
where $M \sim v = 246$~GeV and $\epsilon \sim 0$, corresponding to couplings scaling with masses.
As we see in the right panel of Fig.~\ref{fig:epsilonM1}, the hypothesis $M = v$ is indeed favoured. The left panel of
Fig.~\ref{fig:epsilonM1} tells us that small values of $\epsilon$ are also favoured, and the Higgs hypothesis $\epsilon = 0$
is quite compatible with the available data. Our global fit yields
\beq
\epsilon \; = \; 0.05 \pm 0.08, \; \; M \; = \; 241 \pm 18~{\rm GeV}.
\label{epsilonM}
\eeq
At first sight, one might be surprised that it is already possible to obtain
such a tight constraint on $\epsilon$. The essential reason is that, because it is so much heavier than all the other fermions,
the coupling to the top quark provides a long lever arm, and similarly for the $W^\pm$ and $Z$ because they are also much heavier
than the other fermions.
As already commented in Section~1, Fig.~\ref{fig:Mdep} provides another way of visualizing this observation. 
The diagonal line in Fig.~\ref{fig:Mdep} that represents the mass dependence of the Higgs couplings expected in the Standard Model
(\ref{SMcouplings}) is completely compatible within errors with the measurements.

We consider Figs.~\ref{fig:epsilonM2}, \ref{fig:epsilonM1} and \ref{fig:Mdep} to be the most remarkable results of our analysis.

\section{Overview and Prospects}

A new particle has been discovered: how closely does it resemble the Higgs boson
of the Standard Model? In this paper we have presented a global analysis of the data from the Tevatron
experiments~\cite{TevatronJulySearch} CDF, D0, ATLAS~\cite{ATLASICHEP2012} and CMS~\cite{CMSICHEP2012}
made available before the ICHEP 2012 conference, making two types of fit.
One is in terms of universal coefficients $(a, c)$ that parametrize the deviations of the $h$
couplings to fermions and bosons in a way well adapted to constraining composite Higgs models
such as the pseudo-dilaton, MCHM4 and MCHM5 models, as well as the `anti-dilaton', fermiophobic
and gaugephobic scenarios. As seen in Figs.~\ref{fig:ac2} and \ref{fig:ac1},
the only models favoured in this fit are pseudo-dilaton and MCHM4 models
with parameters close to the Standard Model case. We have also made a fit with $h$ couplings to fermions
and bosons scaling as some power $1 + \epsilon$ of the particle masses, with a normalization scale
$M \ne v$ in general. As seen in Figs.~\ref{fig:epsilonM2}, \ref{fig:epsilonM1} and \ref{fig:Mdep},
this fit favours $\epsilon \sim 0$ and $M \sim v$, as expected for the Standard Model Higgs boson.

\begin{figure}
\vskip 0.5in
\vspace*{-0.75in}
\begin{minipage}{8in}
\hspace*{-0.7in}
\centerline{\includegraphics[height=8cm]{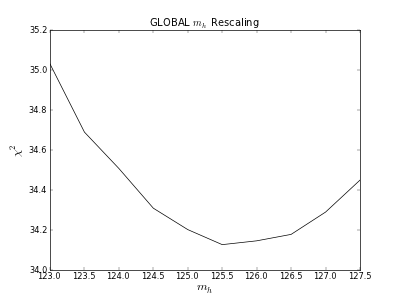}}
\hfill
\end{minipage}
\caption{
{\it
The overall $\chi^2$ of a global fit to the available CDF, D0, ATLAS and CMS data
as a function of $m_h$, obtained by calculating the $h$ production cross-sections
and decay branching rates assuming Standard Model couplings, but not including
information on the shapes of the $h$ signal in, e.g., the high-resolution $\gamma \gamma$ 
and $ZZ^*$ channels as functions of $m_h$.
}} 
\label{fig:mhdep} 
\end{figure}

As seen in Fig.~\ref{fig:mhdep}, the overall quality of a fit to the Standard Model Higgs boson is good,
and does not depend strongly on $m_h$. This plot was made by calculating the $h$ production cross-sections
and decay branching rates assuming Standard Model couplings,
i.e., $a = c = 1$, $\epsilon = 0, M = v$, while leaving $m_h$ as a free parameter. Note that information on
the shapes of the $h$ signal in, e.g., the high-resolution $\gamma \gamma$ and $ZZ^*$ channels as functions of $m_h$
was not used in this exercise. The value of the global $\chi^2$ function at the minimum, namely 34.1, is comparable to
the number of degrees of freedom in the fit. Better understanding of the correlations in the data are needed, but
the overall quality of the Standard Model Higgs fit is clearly good. We also see in Fig.~\ref{fig:mhdep} that the quality of this fit does
not vary significantly over the range $[124, 127]$~GeV, which brackets the central values of $m_h$ found in
the high-resolution $\gamma \gamma$ and $ZZ^*$ channels by CMS and ATLAS. Within this range, our other results are
insensitive to the value $m_h = 125$~GeV assumed in our global fits.

We anticipate that the LHC experiments will be able to constrain the $h$ couplings
significantly further in the coming months, with improved analyses of the channels
already studied, analyses of more channels using the data accumulated so far, and
the prospect of more data on the way. We expect that these improvements will
enable the ranges of parameters in simple two-parameter fits such as those presented here to be reduced
to the 10\% level~\cite{generalcouplings}. The upcoming data should also make possible more detailed fits
incorporating more parameters, leading eventually to individual determinations of
the $h$ couplings to different bosons and fermions. In this way, we shall see whether
the indication of $h$ couplings depending linearly on other particle masses seen
in Figs.~\ref{fig:epsilonM2}, \ref{fig:epsilonM1} and \ref{fig:Mdep} will be confirmed.

So far, the $h$ particle does indeed walk and quack very much like a Higgs boson.

\section*{Acknowledgements}
This work was supported partly by the London
Centre for Terauniverse Studies (LCTS), using funding from the European
Research Council via the Advanced Investigator Grant 267352.

\end{document}